\documentclass[11pt,leqno,twoside]{article}
\usepackage{latexsym,theorem}
\usepackage{amssymb,amsmath}    

\makeatletter

\font\titleft=cmbx12

\def\headlineft{\small\textit}
\def\authorft{\sc}
\def\affiliationft{\sl}
\def\sectionft{\bf}

\usepackage{amsfonts}
\usepackage{amssymb}

\newcommand{\Title}[1]{{\baselineskip=16pt minus 1pt%
 \vskip1in
\begin{center}{\titleft
\uppercase{#1}}\thispagestyle{empty}\end{center}
\vskip\belowdisplayskip}}
\newcommand{\shorttitle}[1]{\def\rightheadline{\hfill\headlineft#1\hfill}}
\def\eet#1{}
\def\supported#1{${^#1}$}

\newcommand{\Author}[1]{
\bgroup\null\let\supported\eet
\global\edef\authorname{#1}\egroup
\bigskip{\baselineskip=14pt
\begin{center}{\authorft#1}\end{center}}
\def\leftheadline{\hfill\headlineft\authorname\hfill}
\markboth{\leftheadline}{\rightheadline}}
\newcommand{\affiliation}[1]{{\baselineskip=14pt\vskip\abovedisplayskip
\begin{center}{\affiliationft #1}\end{center}\vskip\belowdisplayskip}}
\newcommand\dedication[1]{}
\newcommand{\Abstract}[1]{{\footnotesize\noindent #1\par}}
\newcommand{\subject}[1]{\smallskip\noindent{{\footnotesize{\sl AMS subject classifications: }{#1}.\par}}}
\newcommand{\keywords}[1]{{\footnotesize\noindent{\sl Keywords and phrases: }#1.\par}}

{\theorembodyfont{\rmfamily}}

{\theorembodyfont{\rmfamily}\newtheorem{remark}{Remark}[section]}

\newcommand{\email}[1]{{\small {#1}}}

\long\def\endproof{\qed\par\endgroup
        \ifdim\lastskip<\bigskipamount\removelastskip\penalty55\bigskip\fi}
\def\quote{\smallskip\bgroup\advance\leftskip by \parindent}
\def\endquote{\smallskip\egroup\noindent}

\renewcommand{\section}{\@startsection
{section}{0}{0pt}{-\bigskipamount}{\medskipamount}{\sectionft}}
\renewcommand{\subsection}{\@startsection{subsection}{1}{0pt}{-\bigskipamount}%
{\medskipamount}{\sectionft}}%
\newcommand{\acknowledgements}[1]{\bigskip\bigskip\noindent {\bf Acknowledgements. }{#1}}
\def\myref{\smallskip\interlinepenalty 10000
		\hangindent=3pc \hangafter=1 \noindent}

\newcommand{\references}[1]{\let\ref\myref\tolerance=10000
\bigskip\bigskip\begin{center}REFERENCES\end{center}{#1}\bigskip}
\newcommand{\Ref}[1]{\par{\vbox{\hangindent=3pc \hangafter=1 {\noindent {#1}} }\smallskip}}
\newcommand{\address}[1]{\bigskip\bigskip\noindent{\sc {#1}}}

\makeatother

\def\E{\mathord{\rm E}}              
\def\Pr{\mathord{\rm P}}             
\def\Var{\mathop{\rm Var}\nolimits}  

\def\qed{\vrule height4pt width4pt depth0pt}

\def\Var{\mathop{\rm Var}\nolimits}    

\def\longleftsquigglyarrow{{\leftarrow\!\sim}}
\def\longrightsquigglyarrow{{\sim\!\rightarrow}}
\def\lapproxeq{{\lesssim}}

\def\E{\mathord{\rm E}}

\begin{document}

\Title
{Accardi contra Bell (cum mundi):\\ The Impossible Coupling\\  ~  \\ Final version,
18/09/02}
 
\shorttitle
{Accardi contra Bell}

\Author
{Richard~D.~Gill (with an appendix by J.-\AA. Larsson)}

\affiliation
{Mathematical Institute, University Utrecht, and\\
Eurandom, Eindhoven.}

\Abstract
{An experimentally observed violation of Bell's 
inequality is supposed to show the failure of local realism to deal with quantum reality.
However, finite statistics and the time sequential nature of real experiments
still allows a loophole for local realism. We show that the randomised design of the 
Aspect experiment closes this loophole. Our main tool is 
van de Geer's (1995, 2000) 
martingale version of the classical Bernstein (1924) inequality guaranteeing,
at the root $n$ scale, 
a not-heavier-than-Gaussian tail of the distribution of a sum of bounded supermartingale
differences. The results are used to specify a protocol for a public bet between
the author and L. Accardi, who in recent papers (Accardi and Regoli, 2000a,b, 2001);
Accardi, Imafuku and Regoli, 2002) has claimed to have produced
a suite of computer programmes, to be run on a network of computers, 
which will simulate a violation of Bell's inequalites. At a sample size of 
twenty five thousand, both error probabilities are guaranteed smaller than about one in 
a million, provided we adhere to the sequential randomized design while Accardi
aims for the greatest possible violation allowed by quantum mechanics.}

\subject
{60G42, 62M07, 81P68}


\section{Introduction}

This  paper is concerned with a celebrated paradox of quantum mechanics. 
Some keywords and phrases are locality, causality, counterfactuals,
EPR (Einstein--Podolsky--Rosen, 1935) correlations, 
the singlet state, entanglement, Bell's (1964) inequalities,
and the Aspect experiment 
(Aspect et al., 1982a,b). However the point of the paper is that almost
the whole story can be told in 
terms of elementary classical probability and statistics. The only
physics you should believe, is that the right mathematical model for
the periodic, smooth, dependence of a certain correlation coefficient on
a certain angle is given by the appropriate sine curve. It seems to me
that this little example should be in every probability and statistics
course as showing the power of probabilistic reasoning and the
importance of statistics in modern day science
(it is for instance in Williams, 2001, chapter 10). Moreover, there is
growing realisation that quantum physicists are up to interesting things
these days (quantum information, quantum computation, quantum
communication), and growing realisation that these things involve
probability and potentially statistics, and that we should get involved
too. So why not take this as an aperatif, before consulting say
Barndorff-Nielsen, Gill and Jupp (2001) or Gill (2001b) for a survey and
a tutorial respectively, on quantum statistical inference: statistical
inference for data coming from quantum experiments. Gill (2001a)---in
another Festschrift---even introduces quantum asymptotic statistics.

The rest of the paper is structured as follows. Section 2 introduces
the ongoing controversy around the application of Bell's (1964) inequality
to quantum mechanics. The inequality is the elementary
\begin{equation}\label{bineq0}
\Pr\{X_1=Y_2\}~\le~\Pr\{X_1 = Y_1 \}+\Pr\{X_2=Y_1\}+\Pr\{X_2=Y_2\}
\end{equation}
concerning  coincidence probabilities
between four $0/1$-valued random variables. 
Its proof is postponed to Section 4.
Though the inequality itself is trivial, the question of whether or not
it should be applicable to certain real-world experiments is more subtle,
and therein lies the controversy.  The interesting fact is that the inequality is
\emph{apparently} 
violated by experimentally confirmed predictions of quantum mechanics.
 
In Section 3 we describe the celebrated Aspect experiment, which first confirmed
the violation of Bell's inequality, predicted by Bell himself almost
twenty years earlier. In each of a long sequence of runs or trials,
a pair of photons are emitted from a source $O$ and sent to two widely separated
polarization filters $X$, $Y$. In each trial, each filter has one of two possible 
orientations (labelled $1$, $2$, supplied by independent agents $A$ and $B$). 
Each photon either passes or does not pass its filter. We encode this with
a $1$ or a $0$.
We set down some notation and describe the empirical finding of the experiments,
concerning the frequencies of various possible outcomes. We shall work with
absolute frequencies rather than relative frequencies or empirical correlations.
This will lead to a clean mathematical analysis without changing the
conclusions.

Accardi and Regoli (2000a,b, 2001) and Accardi, Imafuku and Regoli (2002)
claim to be able to reproduce these frequencies, replacing the source of the
photons and the two polarization filters by three
computers. A software package can be downloaded from 
{\tt http://volterra.mat.uniroma2.it}, though in my
opinion it does not respect the rules of the game, in particular, that
outcomes are $0/1$ valued, revealed at the proper moments,
and there is no missing data.
The author has publicly challenged Accardi to violate Bell's inequality
in an Aspect-style experiment with a version
of this software which verifiably satisfies these rules. In particular,
the outcomes for each trial must be committed to, 
before each new trial.
The challenge is provisionally accepted subject to
finalising details of the protocol. The bet has been fixed at
$3000$ Euros.  The bet will be settled by an independent jury who are only asked
to verify the one-way connections between the computers,  and to observe if the
empirical correlations violate Bell's 
inequality by a pre-agreed margin. The results of this paper allow the author to
determine a protocol which will be acceptable for him. At the time of writing
negotiations look set to continue indefinitely, Accardi
having stated that my protocol is
``perhaps mathematically interesting but physically irrelevant''.

In the mean time, there have been more challenges to Bell (1964) 
in which an attempt is made to exploit time dependence and memory
effects; see Hess and Philipp (2001a,b). 
I was unable to interest Walter Philipp in a bet: ``our
results are mathematically proven and a computer simulation
is unnecessary''. The present paper provides another mathematically
proved theorem, which contradicts their results, 
see Gill et al.\ (2002), and see Barrett et al.\ (2002) for a related analysis
of the potential memory loophole to Bell's theorem.

In Section 4 we prove
(\ref{bineq0}).  In Section 5 we write down the
probabilities of the events of interest in a version of the Aspect experiment, which
follow from a sine law and certain choices of experimental settings.  It
is not a priori clear that the set-up of Bell's inequality should apply
in the Aspect experiment, but if it did, the results predicted by
quantum mechanics and observed in the physics laboratory would be
impossible. In Section 6 we argue why Bell's inequality \textit{should}
apply to Accardi's experiment, with everyday computers connected so
as to mimic the possible communication lines between the calcium atom and the
polarization filters. Since the behaviour of photons at distant filters cannot
be simulated with classical computers, connected so as to respect the
separation between the filters, it follows that quantum mechanics does
make extraordinary predictions, namely, it predicts phenomena which for 
classical physical systems are impossible. 

In order to test inequalities between expected values, one will in practice compute
averages, and must take account of statistical variability of the outcome. 
Now quantum mechanics predicts the same results whether one does one trial in
each of thousands of laboratories, or does thousands of trials, sequentially, in one laboratory.
In the former case one might be prepared to assume independence from one trial to
another, but in the latter case, it is harder to rule out.  In the case of a computer
network simulation, in which the software has been written by an opponent, one
cannot rule out anything at all.
In Section 7 we show, using the martingale Bernstein inequality of 
van de Geer (1995, 2000), see also Dzhaparidze and van Zanten (2001),
that this does not provide a loophole for the Accardi experiment. 
Twenty five thousand trials carried out according
to a simple protocol are sufficient that both Gill's and Accardi's error probabilities 
are much smaller than one in a million. 

Recent research by this author has shown that
the Hoeffding (1963) inequality, see Bentkus (2002) for the latest improvements,
gives even better results, and this will be reported in a future paper.

Section 8 contains some closing remarks and
further references.

\section{Accardi contra Bell}
\label{s:intro}

Quantum mechanics makes statistical (or if you prefer, probabilistic)
predictions about the world. Some of the strangest are connected to the
phenomenon of \textit{entanglement}, whereby two quite separate quantum
systems (for instance, two distant particles) behave in a coordinated
way which cannot be explained classically. Despite the fact that these
properties are well known and experimentally verified (for instance,
see Tittel et al.\ (1998),
with pairs of photons passing below Lake Geneva
through Swiss Telecom's glass-fibre cable network,
between locations $10$ Km distant from one another) controversy still
surrounds them.

A popular explanation of entanglement runs something like this.  ``Paint
one ping-pong ball red, another blue; put them in closed boxes and send
them randomly to two distant locations.  Before the boxes are opened
either box could contain either ball. If one box is opened and turns out
to contain a red ball, then far away and instantaneously, the state of
the other box suddenly changes: it contains a blue ball.'' This is what
Reinhard Werner calls the ping-pong ball test: to judge any popular
explanation of some quantum mechanical paradox, replace the objects in
the story by ping-pong balls, and check if it makes sense. Well, this
ping-pong story does make sense, but misses the point. The
behaviour we are trying to explain is a bit more complex (too complex
for newspaper articles, but not too complex for mathematical
statisticians). I will describe it precisely in the next section.
Quantum mechanics would not have caused
scientists of the calibre of Schr\"odinger, Bohr, and Einstein such
intellectual discomfort if it were this easy to explain entanglement.
The whole point which Bell was trying to make with his
inequalities is that the
dependence in the behaviour of distant but entangled particles is
contradictory to `local realism'.  Loosely speaking, this phrase means a
classical (though possibly probabilistic) explanation of the correlation
in the behaviour of such particles, through their carrying information
from the place where they were generated or  `born' to the places where they are
measured or observed. In other words, a story like the ping-pong story
will \emph{not} explain it.

Repeatedly, elaborate and exotic theories have been put forward to explain
away the problem. Non-measurable events (Pitowsky, 1989), $p$-adic
probabilities (Khrennikov, 1995a,b, 1997, 1998, 1999), 
and most recently, the chameleon
effect (Accardi et al., 2000a,b, 2001, 2002)
have all been tried. In the mean time much
of the physics community ignores the controversy, and many have
misunderstood  or  minimalised Bell's contribution, which goes
back, via Bohm, to a celebrated thought
experiment of Einstein, Podolsky and Rosen (1935). To give a local example,
Nobel prize-winner G. 't Hooft learnt from his uncle N. van Kampen, 
a  staunch adherant of the Copenhagen interpretation, 
that Bell's inequalities were not worth much attention, 
since they are derived by consideration of what would have
happened if a different experiment had been performed, which according
to Bohr's Copenhagen school is taboo. Counterfactuals have a bad name
in quantum physics. Consequently 't Hooft (1999) was at first
unaware, that a deterministic and classical hidden
layer behind quantum mechanics---such as the one he is attempting to develop
himself---is forced to be grossly non-local.  He now has the onerous task of
explaining why it is that, although every part of the universe is
connected with invisible and instantaneous wiring to every other part,
reality as we know it has that familiar `local' look.

To return to the exotic explanations, Accardi in a number of papers has
strongly argued that the randomness in quantum mechanics is not the
randomness of urns, but of chameleons. By this he means that in
classical probability, with the paradigm being choosing a ball out of an
urn containing balls of different colours, the values of variables on
the different outcomes are fixed in advance. A ball in an urn already
has a particular colour, and this colour is not influenced by taking the
ball out of the urn and looking at it. However the colour of a
chameleon, let loose out of its cage, depends on its environment.
Moreover if there is a chance that the chameleon is mutant, we will not
be able to predict in advance what colour we will see. His image of the
Aspect experiment has a pair of chameleons, one mutant and one normal,
instead of the pair of ping-pong balls. There is some value in this
imagery. Bell's findings reinforce Bohr's philosophy,  that in quantum mechanics 
one should not think of the values of physical quantities 
as being fixed in advance of measurement, and
independently of the total experimental set-up used to elicit the outcomes.
However, in my opinion, if chameleons are to be thought of as classical
physical objects (they may be mutant but not telepathic) it will not be
possible to simulate quantum systems with them. 
But Accardi et al.\ (2000b, 2001, 2002) claim that they have 
simulated Accardi's chameleons 
on a network of PCs. The programme can be downloaded from
{\tt http://volterra.mat.uniroma2.it}.
I have much respect for Accardi's many solid and deep contributions to
quantum probability and quantum physics. On the other hand I cannot find
fault with Bell's argument. I have therefore bet Luigi Accardi $1000$ Euro
(raised at his request to $3000$ in view of the more stringent programming
requirements which I have put down) that he cannot violate Bell's inequalities, 
in an experimental setup to be outlined below. Preparation of this bet
required me to take a new look at the inequalities and in particular to
study the effect of possible time dependence in repeated trials. Most
mathematical treatments consider one trial and then invoke the law of
large numbers and the central limit theorem, assuming independence. Now,
quantum mechanics makes the same predictions when one independently
carries out one trial each in many laboratories over the world, as when
one makes many trials sequentially at one location. Actual
experiments, in particular Accardi's computer experiment, are done
sequentially in time. In order to show that sequentially
designed classical experiments (in particular, using computers or
chameleons) cannot simulate quantum systems, we are not able to assume
independence. It will become clear that it is essential that the
experiment is randomised and the randomization is disclosed sequentially,
with the outcomes of the trials also being disclosed sequentially, in step.
We will see that a martingale structure will prevent the computers from
taking advantage of information gathered in past trials. Put another
way, the separation in time of consecutive trials will play a similar
role to the separation in space which is already central to Bell's
inequality. 

\section{The Aspect experiment} \label{s:experiments}

In an experiment carried out in Orsay, Paris, in 1982 by Alain Aspect
and his coworkers, a calcium atom $O$ is excited by a lazer, and then
returns to its unexcited state by emitting a pair of photons in equal
and opposite directions. The photons
always have equal polarization (in some versions of the experiment,
\emph{opposite} rather than \emph{equal}). In fact, their joint state of
polarization is a so-called quantum entangled state having rather
remarkable properties, as we will see. This is repeated many times
(and there are many calcium atoms involved), 
producing a long sequence of $n$ pairs of photons. We
will refer to the elements of this sequence as `trials'. 

Each pair of photons speed apart until intercepted by a pair of
polarization filters $X$ and $Y$, at two locations several meters apart in the
laboratory. We will call these locations `left' and `right'. The
orientations of the polarization filters can be set, independently at
the two locations, in any desired direction. Aspect wanted that
at each location a series of independent random choices between 
two particular directions was made, independently at the two locations,
and each time in the short time span \emph{while the photons were in flight}.
In 1982 it was not possible to achieve this ideal, and Aspect made
do with a surrogate. We will see that good randomization is absolutely crucial. 
Recent experiments have neglected this, with the notable exception 
of Weihs et al.\ (1998) who could claim
to be the \emph{only} people so far to have actually carried out the
Aspect experiment as Aspect intended; see Aspect (2002).

Each photon either passes or does not pass through its filter. What
happens is registered by a photo-detector. The experiment thus
produces, in total, four sequences of binary outcomes: the filter-settings, 
both left and right, and the outcomes `photon passes' or `photon doesn't pass',
both left and right.

We will be particularly interested in the following event 
which either does or does not happen at each
trial, namely, `the two photons do the same': both pass or neither passes.
Each trial is characterized by one of four possible 
combinations of settings
of the two filters. We label these combinations by
a pair of indices $(i,j)$,  $i=1,2$ for the left setting and $j=1,2$ for the right
setting (we will be specific about the particular orientations
later). Since at each trial, $i$ and $j$ are chosen independently and
with equal probabilities, the four joint outcomes of the settings will
occur approximately equally often, each approximately $n/4$ times. Let
$N_{ij}$ denote the number of times that the two photons do the same,
\textit{within} the subset of trials with joint setting $(i,j)$. In Section 6 
we will argue that in a `local realistic' description of what is
going on here, one will have
\begin{equation}\label{e:bell}
N_{12} ~ \lapproxeq ~  N_{11}+N_{21}+N_{22}.
\end{equation}
In fact one has four inequalites: each
of the four random counts should be less than the sum of the other
three, modulo random noise, which is what we indicate with the 
`approximate inequality' symbol.
Violation of the inequality, if at all, would be due to
statistical variation and therefore at most of the order of $\sqrt n$,
if one may assume independence between the trials.
If we allow for sequential dependence then perhaps a worse violation
could occur by chance, and it is the purpose of this
paper precisely to quantify how large it could be. 

Quantum mechanics predicts that, if the angles are chosen
suitably, one can have
\begin{equation}\label{e:qm}
N_{12} ~ \gg ~ N_{11}+N_{21}+N_{22},
\end{equation}
and this is what Aspect et al.\ (1982a,b) experimentally verified; in
particular the second paper introduced the randomly varying polarization
filter settings. Nowadays this experiment can be done in any decent
university physics laboratory, though twenty years ago the experiment
was a tour de force. In fact one usually replaces the absolute frequencies
in the equations (\ref{e:bell}) and (\ref{e:qm}) by relative frequencies.
Since the denominators will be roughly equal, this does not make much
difference, and working with absolute frequencies 
allows a much cleaner mathematical analysis below.

Actually I am simplifying somewhat and will not go into the major
complications involved when one takes account of the fact that not all
emitted photons are detected. To be honest it must be said that this
still leaves a tiny, but rapidly dissappearing, loophole for local
realism in ever more conspiratorial forms.
For the latest theoretical progress in this area see Larsson (2002), 
Larsson and Semitecolos (2001), Massar (2001); and for experimental progress,  
Weihs et al.\ (1998), Rowe et al.\  (2001).

Accardi et al.\ (2000b, 2001, 2002) claim that they can programme three
computers, one representing the calcium atoms and sending information to
two other computers, representing the polarization filters, to reproduce
the predictions of quantum mechanics, or at the least, to satisfy
(\ref{e:qm}). My bet is that their experiment will however reproduce
(\ref{e:bell}). My protocol of the experiment stipulates that I provide
two streams of binary outcomes to each of the two `polarization
filters', representing the choices of setting (orientation) of each
filter. Graphically one trial of the experiment looks like this:
\begin{equation}
A \quad\longrightarrow\quad X\quad \longleftsquigglyarrow\quad O 
\quad\longrightsquigglyarrow \quad Y \quad \longleftarrow \quad B
\end{equation}
\begin{equation*}
\downarrow\qquad \qquad \downarrow \qquad      \qquad \qquad  
 \qquad \downarrow \qquad \qquad\downarrow
\end{equation*}
where $X$ and $Y$ denote the two polarization filters, $O$ denotes the
calcium atom, $A$ and $B$ are two operators (Alice, Bob) independently
choosing the settings at $X$ and $Y$. The downwards arrows coming from
$A$ and $B$ represent exact copies of the settings sent by $A$ and $B$
to $X$ and $Y$. The wiggly arrows emanating from $O$ are supposed
to suggest a quantum rather than a classical (straight) connection.
Accardi claims he can replace them with straight arrows. The statistician
must process four downward streams of binary data: the settings
from $A$ and $B$, and the outcomes from $X$ and $Y$.

\section{Bells' inequality}
\label{s:bell}
This little section derives Bell's inequality, which lies behind the
prediction (\ref{e:bell}). For the time being treat this as a background
fact from probability theory. Why it should be relevant to a local
realistic version of the Aspect experiment, we will argue in Section 6.
Actually, the inequality I prove is a form of the ``CHSH'', i.e.,
Clauser--Horne--Shimony--Holt (1969) version of
Bell's inequality,
better tuned to a stringent experimental distinction between quantum
mechanical and classical systems. The way it will be proved here, as a probabilistic
consequence of a deterministic inequality, is often attributed to Hardy (1993).
In fact, others also earlier used this arguement, and its seeds are already in Bell's
paper.  Some trace the inequality back to the works of the nineteenth century logician Boole.
I learnt it from Maassen and K\"ummerer (1998). Bell himself,
along with most physicists, gives a more involved proof, since the physics
community does not make use of standard probabilistic notation 
and arguments. I also prefer, for
transparency, an inequality in terms of probabilities of coincidences to
one in terms of correlations (which however are what the physicists
prefer to talk about).

Let $X_1$, $Y_1$, $X_2$, $Y_2$ denote four $0/1$-valued random
variables. Think of them positioned at the vertices of a square, with
$X_1$ opposite to $X_2$, $Y_1$ opposite to $Y_2$. Each side of the
square connects one of the $X$ variables to one of the $Y$ variables,
and therefore represents an experiment one could possibly do with two
photons and two polarization filter settings. Convince yourself, by
following through the choice of a $0$ or a $1$ for $X_1$, that
\begin{equation}
X_1 \ne Y_1 ~\&~ Y_1 \ne X_2 ~\&~ X_2 \ne Y_2~~ \implies~~ Y_2 \ne X_1 .
\end{equation}
Taking the negation of each side and reversing the implication, it follows that
\begin{equation}
X_1=Y_2~~ \implies ~~X_1 = Y_1~ \mbox{or} ~ X_2=Y_1~ \mbox{or}~ X_2=Y_2 .
\end{equation}
Now use one of the first properties of probability:
\begin{equation}\label{e:bellxy}
\Pr\{X_1=Y_2\}~\le~\Pr\{X_1 = Y_1 \}+\Pr\{X_2=Y_1\}+\Pr\{X_2=Y_2\} .
\end{equation}

If you are interested in correlations, by which the physicists mean 
\emph{raw product moments}, note that 
(physicist's notation) $\langle X_1,Y_2\rangle=\E(X_1 Y_2)= 2\Pr\{X_1=Y_2\}-1$.

\section{Coincidence probabilities for entangled photons}

The two photons in the Aspect experiment have in some sense exactly equal
polarization. If the two polarization filters left and right are in
perpendicular orientations, \textit{exactly one} of the two photons will
pass through the filter. For instance, if one filter is oriented horizontally, and
the other vertically, one might imagine that the calcium atom either produces two
horizontally polarized photons, or two vertically polarized photons, each with
probability half. With probability half, both photons are horizontally polarized, and
the one which meets the horizontal filter, passes through it, while the other meets
a vertical filter and is absorbed. With probability half both photons are vertically
polarized and again, exactly one passes the two filters.
The same holds for any two perpendicularly
oriented  filters:
the probability of coincidence---the two photons do the same---is zero. 
If however the two filters are
oriented in the same direction, for instance, both horizontal,
then \textit{either} both photons pass, or both do not
pass (each of these possibilities has probability half). The probability
of coincidence is one. Now imagine keeping one filter fixed and slowly
rotating the other. At zero degrees difference, the probability of
coincidence is $1$, at $90$ degrees, it is $0$, at $180$ degrees it is
back to one, and so on. It is a smooth curve (how could it not be
smooth?), varying periodically between the values $0$ and $1$. Recalling
that $\cos(2\theta)=2\cos^2(\theta)-1$, and that the cosine function is itself
a shifted sine curve, we conclude that if the probability of
coincidence is a sine curve, it has to be the curve $\cos^2(\theta)$:
it varies between $0$ and $1$, taking these values at $\theta=\pi/2$
and $\theta=0$.

Quantum mechanics predicts precisely this probability of coincidence
(in fact so does classical optics, but there light comes in continuous
waves, not discrete particles, and the word ``probability'' has to be replaced
by ``intensity'').
The quantum state involved, is the only pure state having the natural rotational
invariance so this answer is pretty canonical. Recall that quantum
mechanics is characterized by wave-particle duality: we know that
photons are particles, when we look to see with a photo-dector if one is
present or not. But we also know that light behaves like waves,
exhibiting interference patterns. Waves are smooth but particles,
especially deterministic particles, are discrete. However, random
particles can have smoothly varying behaviour. It seems that randomness
is a necessary consequence of the fundamental
wave-particle duality of quantum mechanics, i.e., of reality.

Now suppose $A$ chooses, for $X$, between the orientations $\alpha_1=0$
and $\alpha_2=\pi/3$, while $B$ chooses, for $Y$, between the
orientations $\beta_1=-\pi/3$ and $\beta_2=0$. The absolute
difference between each $\alpha$ and each $\beta$ is $0$, $\pi/3$, 
or $2\pi/3=\pi-\pi/3$. Since
$\cos(\pi/3)=1/2=-\cos(\pi-\pi/3)$ the four probabilities of coincidence are $1/4$,
$1/4$, $1/4$, $1$, and
\begin{equation}
1 ~\gg~\frac14 +\frac14+\frac14.
\end{equation}
Even better angles are $\alpha_1=\pi/8$,
$\alpha_2=3\pi/8$, $\beta_1=-\pi/4$ and $\beta_2=0$ giving probabilities
of coincidence approximately $0.15$, $0.15$, $0.15$ and $0.85$ and a difference
of $\sqrt 2 -1\approx 4/10$ instead of our $1/4$.

\section{Why Bell applies to Accardi's computers}

Consider one trial. Suppose the computer $O$ sends some information to
$X$ and $Y$. It may as well send the same information to both (sending more,
does not hurt). Call the information $\lambda$. Operator $A$ sends
$\alpha_1$ or $\alpha_2$ to $X$. Computer $X$ now has to do a
computation, and output either a $0$ or a $1$ (`doesn't pass', `does
pass'). In our imagination we can perfectly clone a classical computer:
i.e., put next to it, precisely the same apparatus with precisely the
same memory contents, same contents of the hard disk. We can send
$\alpha_1$ to one of the copies and $\alpha_2$ to the other copy; we can
send $\lambda$ to both (classical information can be cloned too). By the
way, quantum systems cannot be cloned---that is a theorem of quantum
mechanics! Therefore both copies of the computer $X$ can do their work
on both possible inputs from $A$, and the same input from $O$, and
produce both the possible outputs. Similarly for $Y$.

Let us now suppose that this is actually the $m$th trial. I allow that
computers $O$, $X$ and $Y$ use pseudo-random number generators
and that I model the seeds of the generators with random variables. This
means that I now have defined four random variables $X_{m1}$, $X_{m2}$,
$Y_{m1}$ and $Y_{m2}$, the values of two of which are actually put on
record, while the other two are purely products of your and my
imagination. \textit{Which} are put on record is determined by independent
(of everything so far) Bernoulli trials, the choices of $A$ between
index $1$ or $2$ for the $X$ variables, and of $B$ between index $1$ or
$2$ for the $Y$ variables. Let me directly define variables $U_{m11}$,
$U_{m12}$, $U_{m21}$, $U_{m22}$ which are indicator variables of the
four possible joint outcomes. Thus the sum of these four $0/1$ variables
is identically $1$, and each is Bernoulli$(\frac14)$. I will allow
Accardi's computers, at the $m$th trial, to use results obtained so far
in its computations for the current trial. So we arrive at the following
model: for each $m=1,...,n$, the vector
$(U_{m11},U_{m12},U_{m21},U_{m22})$ is
multinomial$(1;\frac14,\frac14,\frac14,\frac14)$, independent of all
preceding $U$, $X$ and $Y$ variables, and independent of the current $X$
and $Y$ variables. The counts on which the bet depends are
$N_{ij}=\sum_m U_{mij}1\{X_{mi}=Y_{mj}\}$. I compute the expectation of
this by first conditioning, within the $m$th term, on the current and
preceding $X$ and $Y$ variables and on the preceding $U$ variables. By
conditional independence and by taking the expectation of a conditional
expectation I find $\E N_{ij}=\frac14\sum_m \Pr\{X_{mi}=Y_{mj}\}$.
Therefore
\begin{equation}
\E(N_{12}-N_{11}-N_{21}-N_{22})
\end{equation}
\begin{equation*}=\frac14\sum_m
\Bigl( \Pr\{X_{m1}=Y_{m2}\}-\Pr\{X_{m1} = Y_{m1}\}-
\Pr\{X_{m2}=Y_{m1}\}-\Pr\{X_{m2}=Y_{m2}\}
\Bigr)
\end{equation*}
\begin{equation*}
 ~\le ~ 0,
\end{equation*}
by Bell's inequality (\ref{e:bellxy}). \emph{In expectation} Accardi
must lose. If each trial is independent of each
other, the deviation can be at most of the order of $\sqrt n$.
In
the next section we will see that serial dependence cannot worsen this
at all, because of the obvious (super)martingale structure in the variable
of interest.

\section{Supermartingales}

Let us allow the choices of computers $O$, $X$ and $Y$ at the $m$th
trial to depend arbitrarily on the past up to that time. Write $\vec
1=(1,1,1,1)$, 
\begin{equation*}
\vec
U_m=(U_{m1},U_{m2},U_{m3},U_{m4})=(U_{m12},U_{m11},U_{m12},U_{m22})
\end
{equation*} 
and
\begin{equation*}
\vec X_m=(1\{X_{m1}=Y_{m2}\},
-1\{X_{m1}=Y_{m1}\}, -1\{X_{m2}=Y_{m2}\},
-1\{X_{m1}=Y_{m2}\}).
\end{equation*}
Define $\Delta_m=\vec U_m\cdot
\vec X_m$ and $S_m=\sum_{m=1}^r \Delta_r$.
Define $\Delta_m^\ast=\frac14\vec 1\cdot \vec X_m$.
Let $\mathcal F_m$ denote the
$\sigma$-algebra of all $X$, $Y$ and $U$ variables up to and including
the $m$th trial. 
Let the smaller $\sigma$-algebra $\mathcal A_m$ be the
$\sigma$-algebra generated by $\mathcal F_{m-1}$ together with $X_{m1}$,
$X_{m2}$, $Y_{m1}$, $Y_{m2}$. Thus $\mathcal F_m$ is generated by
$\mathcal A_m$ together with $\vec U_m$. 
Define $\widetilde\Delta_m=\E(\Delta_m|\mathcal F_{m-1})$
and $\widetilde S_m=\sum_{r=1}^m \widetilde\Delta_r$.
In the previous section we
basically  made the computation 
$\widetilde\Delta_m=\E(\Delta_m|\mathcal
F_{m-1})=
\E(\E(\Delta_m|\mathcal A_m) | \mathcal
F_{m-1}))=\E(\Delta_m^*|\mathcal F_{m-1})$ where
surely, $-1/2 \le\Delta_m^* \le 0$, therefore also
 $-1/2 \le \widetilde \Delta_m\le 0$ and 
$|\Delta_m-\widetilde \Delta_m|\le 3/2$. Define
$\sigma^2_m=\Var(\Delta_m-\widetilde \Delta_m | \mathcal F_{m-1})$. 
Using the facts that the support of $\Delta_m$ is $\{-1,0,1\}$ with
probabilities of the extreme values bounded by $3/4$ and $1/4$
one easily finds $0\le \sigma_m^2\le \frac 34$ almost surely.
Define 
$V_m=\sum_{r=1}^m \sigma^2_r$.

To warm up, we investigate whether we can obtain a Chebyshev-like
inequality in this situation. The answer will be yes, but the
inequality will be too poor for practical use. After that we will
make better use of the fact that all summands are bounded, and
derive a powerful Bernstein-like inequality.

It follows from the computations above that $S_m-\widetilde S_m$ is a
martingale with respect to the filtration $(\mathcal F_m)_{m=1}^n$, and
so is $(S_m-\widetilde S_m)^2-V_m$, while $\widetilde S_m$ is a
decreasing, negative, predictable process and $V_m$ an increasing,
positive, predictable process. By the inequality of Lenglart (1977) it
follows that for any $\eta>0$ and $\delta>0$,
$\Pr\{\sup_{m\le n}(S_m-\widetilde
S_m)^2\ge \eta\}~\le~\delta/\eta+\Pr\{V_n\ge\delta\}$. Choosing
$\eta=k^2 n$ and noting that $V_n\le 3n/4$, we find the inequality
\begin{equation}
\Pr\{S_n\ge k\sqrt n\}~\le~\frac \delta{k^2 n}
+\Pr\{V_n\ge\delta\}~\le~\frac \delta{k^2 n}+\frac {3n}
{4\delta},
\end{equation}
by Chebyshev's inequality. The right hand side
is minimal at $\delta=\sqrt 3 n/2k$ giving us the
inequality
\begin{equation}
\Pr\{S_n\ge k\sqrt n\}~\le~\frac {\sqrt 3}k .
\end{equation}
This is nowhere as good as the result of applying
Chebyshev's inequality when all trials are
independent,
\begin{equation}
\Pr\{S_n\ge k\sqrt n\}~\le~\frac 1{k^2},
\end{equation}
but it would allow us to choose a (huge)
sample size and critical value to settle my bet with Luigi Accardi. Note
that I can for free replace $S_n$ by $\sup_{m\le n} S_m$ in these
inequalities, so there is no chance that Accardi can win by stopping
when things are looking favourable for him (they
won't).  However the sample size is prohibitively large, for the rather 
small error probabilities which we would like to guarantee.

In fact we can do much better, using exponential bounds for martingales,
generalizing the well-known Bernstein (1924), Hoeffding (1963), or Bennett (1962)
inequalities
for sums of bounded, independent random variables, and more generally for independent random 
variables with bounded exponential moment. From 
van de Geer (1995), or van de Geer (2000, Lemma 8.11),
applied to the martingale $S_m-\widetilde S_m$ whose 
differences are bounded in absolute value by $3/2$ with conditional variances
bounded by $3/4$
we obtain:
\begin{equation}\label{bernstein}
\Pr\{\sup_{m\le n} S_m \ge \frac {\sqrt 3}2 k\sqrt n\}~\le~
\exp\Biggl(-\frac12 k^2\Bigl(\frac 1 {1+\frac 1{\sqrt 3} \frac k{\sqrt n}}\Bigr)\Biggr).
\end{equation}
More precisely, van de Geer (1995, 2000) gives us the stronger result obtained by replacing
$S_m$ with $S_m-\widetilde S_m$ in (\ref{bernstein}), but we also know that 
$\widetilde S_m\le 0$.
Thus at the root $n$ scale, the tail of our statistic
can be no heavier than Gaussian; though for much larger values
(at the scale of $n$) it can be as heavy as exponential. This behaviour
is no worse than for sums of independent random variables. In fact if $S_n$ 
denotes the sum of $n$ independent random variables each with mean zero, 
bounded from above by $3/2$, and variance bounded by $3/4$,
the classic Bernstein inequality is simply
(\ref{bernstein}) with $\sup_{m\le n} S_m$ replaced by $S_n$.
The discrete time martingale maximal Bernstein inequality goes back to 
Steiger (1969) and Freedman (1975), while Hoeffding (1963) already had a
martingale maximal version of his, related, exponential inequality.
A continuous time martingale version of the Bernstein inequality
can be found in Shorack and Wellner (1986). A recent treatment of the inequality for
independent random variables can be found in Pollard (2001, Ch.~11).

Note that if we had been working with the relative instead of the absolute
frequencies, we could have treated the four denominators in the same way,
used Bonferroni, and finished with a very similar but messier inequality.

We can now specify precisely a protocol for the computer experiment, which must settle
the bet between Accardi and the author. In order that the supermartingale structure is present,
it suffices that the settings and the outcomes are generated sequentially: Gill provides settings
for trial 1, then Accardi provides outcomes for trial 1, then Gill provides settings for trial 2, 
Accardi outcomes for trial 2, and so on. Between subsequent trials, computers $X$, $O$ and $Y$
may communicate with one another in any way they like. Within each trial, the communications
are one way only, from $O$ to $X$ and from $O$ to $Y$; and from $A$ to $X$ and from $B$ to $Y$.
A very rough calculation from (\ref{bernstein}) shows that if both accept error probabilities
of one in a million, Accardi and Gill could agree to
a sample size of sixty five thousand, and a critical value $+n/32$,
half way between the Bell expectation bound $0$ and the Aspect experiment 
expectation $+n/16$. I am supposing here that Accardi plans not just to violate the
Bell inequality, but to simulate the Aspect experiment with the filter settings 
as specified by me. I am also supposing that he is happy to rely on Bernstein's inequality, in the
opposite direction. Only twenty five thousand trials are needed 
when Accardi aims for the greatest violation allowed under quantum mechanics, namely 
an expectation value of approximately $+n/10$ and critical value $+n/20$.

The experiment will be a bit easier to perform, if Accardi does not want to exploit the
allowed communication
between his computers, \textit{between} trials. 
In that case one might as well store the entire 
initial contents of memory and hard disk, of computer $O$, within computers $X$ and $Y$.
Now computers $X$ and $Y$ can each simulate computer $O$, without communicating with one
another.
Now we just have computers $A$ and $X$, connected one-way, and completely separately,
$B$ and $Y$, also connected one-way. We carry out $n$ sequential
trials on each pair of computers.

It would be even more convenient if these trials could be 
done simultaneously, instead of sequentially. Thus computers $A$ and
$B$ would deliver to $X$ and $Y$, in one go, all the settings for the $n$ trials. 
We now lose the martingale structure. For the $m$th trial, one can condition on 
all preceding \textit{and} subsequent settings. Conditioning also on the intial contents of
computers $X$ and $Y$, we see that the outcomes of the $m$th trial are now deterministic
functions of the \textit{random} settings for the $m$th trial. Thus we still have Bell's
inequality: in expectation, nothing has changed. But the martingale structure is 
destroyed; instead, we have something like a Markov field. 
Is there still a Bernstein-like inequality for this situation?
It is not even clear if a Chebyshev inequality is available, in view of the possible correlations
which now exist between different outcomes.
However, since we have the Bell inequality in expectation, one could
put the onus on keeping the variance small, on the person who claims
they can simulate quantum mechanical correlations on a classical computer.
For instance, Accardi might believe that he can keep the second decimal
digit of $N_{ij}/n$ fixed, when $n$ is as large as, say, ten thousand.
Then one could do the experiment in ten times four batches of ten thousand,
sending files by internet forty times. Within each group of four batches,
I supply a random permutation of the four joint settings $(i,j)$. 
We settle on a critical value halfway between our two expectations, but
Accardi must also agree to lose, if the second decimal digits of
each group of $10$ $N_{ij}/n$, $n$ being the size of the batch now,
ever vary. Am I safe? I feel uneasy, without Bernstein behind me.

In the actual Aspect experiment,  the alternative set of angles
mentioned above are used, so as to achieve, by an inequality of Cirel'son (1980),
the most
extreme violation of the Bell inequality which is allowed within quantum 
mechanics. Thus if an even larger violation had been observed, one would
not just have had to reject the specific quantum mechanical calculations
for this particular experiment, but more radically have to reject 
the accepted rules of quantum mechanics, altogether. Many authors have
therefore considered those settings as providing ``the most strong
violation of local realism, possible''. However, we would say that 
the strongest violation occurs, when one is able to reject local realism,
with the smallest possible number of samples. Thus concepts of efficiency
in statistical testing, should determine ``the strongest experiment''.

Van Dam, Gill and Grunwald (2002) study this problem from a game-theoretic 
point of view,
in which the believer in quantum mechanics needs to find the experimental
set-up which provides the maximal ``minimum Kullback-Leibler distance between
the quantum mechanical predictions and any possible prediction subject to
local realism''. Such results can be reformulated in terms of size, power,
and sample size, using Bahadur efficiency (large deviations).

Many authors discuss the Aspect experiment and Bell inequalities in a version
appropriate for spin half particles (for instance, electrons) rather than photons.
The translation from photons to electrons is: double the angles, and then
rotate the settings in one wing of the experiment by $180^\circ$. To 
explain the doubling:
a polarization filter behaves oppositely after rotating $90^\circ$, and identically after
rotating $180^\circ$. A Stern-Gerlach magnet behaves oppositely 
after rotating $180^\circ$, identically
after rotating $360^\circ$. As for the rotation: the photons in 
our version of the Aspect experiment are identically
polarized while the spin of the spin half particles in the companion experiment are 
equal and opposite. The quantum state used in the spin half version is the famous
Bell or singlet state, $|01\rangle-|10\rangle$, while for photons one uses
the state $|00\rangle+|11\rangle$, where the $0$ and $1$ stands for ``spin-up'',
``spin-down'' for electrons, and ``horizontal polarization'', ``vertical polarization''  for photons. There are also photon experiments with oppositely polarized rather
than equally polarized photons, and the state $|01\rangle+|10\rangle$.

\section{A different kind of probability, or nonlocality?}

The relation between 
classical and quantum probability and statistics has been a matter of 
heated controversy ever since the discovery of quantum mechanics. 
It has mathematical, physical, and philosophical 
ingredients and much confusion, if not controversy, has been generated by 
problems of interdisciplinary communication between mathematicians, 
physicists, philosophers and more recently statisticians. 
Authorities from both physics and mathematics, 
perhaps starting with Feynman (1951),
have promoted vigorously the standpoint that `quantum probability' 
is something very different from `classical probability'.
Most recently, Accardi and Regoli (2000a)
state ``the real origin of 
the Bell's inequality is the assumption of the applicability of 
classical (Kolmogorovian) probability to quantum mechanics'' which 
can only be interpreted as a categorical statement that classical probability 
is \textit{not} applicable to quantum mechanics.
Accardi et al.'s (2002) aim is ``to show that Bell's statement \dots\  is
theoretically and experimentally unjustified'', and they diagnose Bell's
error as an incorrect use of Kolmogorov probability and conditioning.
Malley and Hornstein (1993) conclude from the perceived conflict between 
classical and quantum probability that 
`quantum statistics' should be set apart from classical statistics.

We disagree. In our opinion, though fascinating mathematical facts and physical 
phenomena lie at the root of these statements, cultural preconceptions
have also played a role. Probabilistic and statistical problems from 
quantum mechanics fall definitely in the framework of classical 
probability and statistics,
and the claimed distinctions have retarded the adoption of 
statistical science in physics. The phenomenon of quantum entanglement
in fact has far-reaching technological implications, which can only be
expressed in terms of classical 
probability; their development will surely involve classical statistics too. 
Emerging quantum technology (entanglement-assisted 
communication, quantum computation, quantum holography and 
tomography of instruments) aims to capitalise on precisely 
those features of quantum mechanics which in the past have often been seen
as paradoxical theoretical nuisances.

Our stance is that the predictions which quantum mechanics makes of the 
real world are stochastic in nature. A quantum physical model of a 
particular phenomenon allows one to compute probabilities of all possible 
outcomes of all possible measurements of the quantum system. The word 
`probability' means here: relative frequency in many independent repetitions.
The word `measurement' is meant in the broad sense of: macroscopic results of
interactions of the quantum system under study with the outside world.
These predictions depend on a summary of the state of the quantum system.
The word `state' might suggest some fundamental property of a particular 
collection of particles, but for our purposes all we need to understand 
under the word is: a convenient mathematical encapsulation of
the information needed to make any such predictions. 

Now, at this formal level
one can see analogies between the mathematics of quantum states and 
observables---the physical quantities of quantum mechanics---on the one hand, 
and classical probability measures and random variables 
on the other. This analogy is very strong and indeed mathematically very 
fruitful (also very fruitful for mathematical physics). Note that 
collections of both random variables 
and operators can be endowed with algebraic structure (sums, products, 
\dots). It is a fact that from an abstract point of view a basic structure 
in probability theory---a collection of random variables $X$ on a countably 
generated probability space, together with their expectations $\int X 
\mathrm d P$
under a given probability measure $P$---can be represented by a 
(commuting)
subset of the set of self-adjoint operators $Q$ on a separable Hilbert space 
together with the expectations $\mathrm{tr}\{\rho Q\}$ computed using the trace 
rule under a given state $\rho$, mathematically represented by another 
self-adjoint operator
having some special properties (non-negative and trace $1$). 

`Quantum probability', or `noncommutative 
probability theory' is the name of the branch of mathematics which studies 
the mathematical structure of states and observables in quantum mechanics.
It is a fact that a \textit{basic} structure in classical 
probability theory is isomorphic to a \textit{special case} of a basic structure in 
quantum probability. 
Brief introductions, of a somewhat ambivalent nature, can be found in the
textbooks, on classical probability, of Whittle (1970) and Williams (2001).
K\"ummerer and Maassen (1998), discussed in Gill (1998), use the 
``quantum probabilistic modelling'' of the Aspect experiment---which just
involves some simple linear algebra involving $2\times 2$ complex matrices---to
introduce the mathematical framework of quantum probability, giving the violation
of the Bell inequalities as a motivation for needing ``a different probability 
theory''.  From a mathematical point of view, one may justly 
claim that classical probability is a special case 
of quantum probability. The claim does entail, however, a rather narrow view 
of classical probability. Moreover, many probabilists will feel that
abandoning commutativity is throwing away the baby with the 
bathwater, since this broader mathematical structure has no analogue of the
sample outcome $\omega$, and hence no opportunity for a
probabilist's beloved probabilistic arguments.

Many authors have taken the probabilistic predictions of quantum theory,
as exemplified by those of the Aspect experiment, as 
a defect of classical probability theory and there have been proposals to 
abandon classical probability in favour of exotic
alternative theories (negative, 
complex or $p$-adic probabilities; nonmeasurable events; noncommutative 
probability; \dots) in order to `resolve the paradox'. However in our 
opinion, the phenomena are real and the defect, if any, lies in 
believing that quantum phenomena do not contradict \textit{classical, deterministic,
physical thinking}. This opinion is supported by the recent development of 
(potential) technology which acknowledges the extraordinary nature of the 
predictions and exploits the discovered phenomena (teleportation, 
entanglement-assisted communication, and so on). In other words, one should 
not try to explain away the strange features of quantum mechanics as some 
kind of defect of classical probabilistic thinking, but one should use 
classical probabilistic thinking to pinpoint these features.

The violation of the Bell inequalities show that any deterministic, underlying,
theory intending to explain the surface randomness of quantum physical predictions,
has to be grossly non-local in character. For some philosophers of science, for
instance Maudlin (1994), this is enough to conclude that ``locality
is violated, \textit{tout court}''. He goes on to analyse, with great clarity, 
precisely what kind of locality is violated, and he investigates possible conflicts with 
relativity theory.
Whether or not one says that locality is violated, depends on the meaning of the word 
``local''. In our opinion, it can only be given a meaning relative to some 
model of the physical world, whether it be implicit or explicit, primitive or sophisticated.

Since quantum randomness is possibly the only \textit{real} 
randomness in the world---all other chance mechanisms, like tossing dice or coins, can
be well understood in terms of classical deterministic physics---there is
justification in concluding that ``quantum probability is a different kind
of probability''. And all the more worth studying, with classical statistical
and probabilistic tools, for that.

\acknowledgements
{I am grateful to Hermann Thorisson for the subtitle of this paper;
see Thorisson (2000) for the connection with the probabilistic
notion of coupling.}


\appendix
\section{What went wrong?}

This appendix is provided by Jan-\AA ke Larsson ({\tt jalar@mai.liu.se}), 
Link\"oping, Sweden. It points out the error in the Accardi and Regoli 
construction. 

In Accardi and Regoli (2001), it is argued that the Bell inequality can be
violated by a classical system after a local dynamical evolution.
After a dynamical evolution, in the Schr\"odinger picture an expectation
is obtained as
\begin{equation}
  E(F)=\iint F(\lambda_1,\lambda_2) (\psi_0\circ P)(d\lambda_1,d\lambda_2),
\end{equation}
while in the Heisenberg picture,
\begin{equation}
  E(F)=\iint P(F)(\lambda_1,\lambda_2) \psi_0(d\lambda_1,d\lambda_2).
\end{equation}
Perhaps it should be underlined here that the two above expressions
are \emph{equivalent representations of the same physical system}.
This means among other things that the possible values of the
observables (values of the random variables, outcomes of the
experiment) in the right-hand sides should be equal, regardless of the
representation. In mathematical language, $R(F)=R(P(F))$.

In Accardi and Regoli (2001), it is claimed that $P(F)$ in the Heisenberg picture
is of a certain form:
\begin{equation}
  P(F)(\lambda_1,\lambda_2)=F(T_{1,a}\lambda_1,T_{2,b}\lambda_2)
  T'_{1,a}(\lambda_1)T'_{2,b}(\lambda_2)\label{eq:claim}
\end{equation}
For the physical system in question in Accardi and Regoli,
$R(F)=\{\pm1\}$, so the only $T_i$s that can be used if
(\ref{eq:claim}) holds are those for which
\begin{equation}
  T'_{1,a}(\lambda_1)T'_{2,b}(\lambda_2)=1\text{\quad a.e.}
\end{equation}
The model (18) in Accardi and Regoli (2001) does not follow this requirement, but
instead, the measurement results in the Heisenberg picture lie in the
interval $[-\sqrt{2\pi},\sqrt{2\pi}]$. The Bell inequality (the CHSH
inequality) is only valid for systems for which the results are in
$\{\pm1\}$ ($[-1,1]$), and for such systems, the correlation is less
exciting.


\references

\raggedright

\Ref{
Accardi, L. and Regoli, M. (2000a).
Locality and Bell's inequality.
Preprint 399, Volterra Institute, University of Rome II.
{\tt http://arXiv.org/abs/quant-ph/0007005}
}

\Ref{
Accardi, L. and Regoli, M. (2000b).
Non-locality and quantum theory: new experimental 
evidence.
Preprint, Volterra Institute, University of Rome II.
{\tt http://arXiv.org/abs/quant-ph/0007019}
}

\Ref{
Accardi, L. and Regoli, M. (2001).
The EPR correlations and the chameleon effect.
Preprint, Volterra Institute, University of Rome II.
{\tt http://arXiv.org/abs/quant-ph/0110086}
}

\Ref{
Accardi, L., Imafuku, K. and Regoli, M. (2002).
On the EPR-chameleon experiment.
{\it Infinite Dimensional Analysis, Quantum Probability and
Related Fields \bf 5}, 1--20.
}

\Ref{
Aspect, A., Dalibard, J. and Roger, G. (1982a).
Experimental realization of {E}instein--{P}odolsky--{R}osen--{B}ohm 
{G}edankenexperiment: a new violation of {B}ell's inequalities.
{\it Phys.\ Rev.\ Letters \bf 49},
91--94.
}

\Ref{
Aspect, A., Dalibard, J. and Roger, G. (1982b).
Experimental test of {B}ell's inequalities using time-varying analysers.
{\it Phys.\ Rev.\ Letters \bf 49},
1804--1807.
}

\Ref{
Aspect, A. (2002).
Bell's theorem: the naive view of an experimentalist.
In: Bertlmann, R.A. and
Zeilinger, A., eds., 
{\it Quantum [Un]speakables. From Bell to Quantum
Information}.
Springer-Verlag, New York.
}

\Ref{
Barndorff-Nielsen, O. E. , Gill, R. D. and Jupp, P. E. (2001).
On Quantum Statistical Inference.
Submitted to {\it J. Roy.\ Statist.\ Soc.\ B}. 
Preprint 2001-19
at {\tt http://www.maphysto.dk}.
}

\Ref{
 Barrett, J., Collins, D., Hardy, L., Kent, A. and Popescu, S. (2002).
Quantum nonlocality, Bell inequalities and the memory loophole.
To appear in {\it Phys. Rev. A}.
{\tt http://arXiv.org/abs/quant-ph/0205016}
}

\Ref{
Bell, J. S. (1964).
On the {E}instein {P}odolsky {R}osen paradox.
{\it Physics \bf 1},
195--200.
}

\Ref{
Bennett, G. (1962).
Probability inequalities for sums of independent random variables.
{\it J. Amer.\ Statist.\ Assoc.\ \bf 57}, 33--45.
}

\Ref{
Bentkus, V. (2002).
On Hoeffding's inequalities.
{\it Ann.\ Probab.}, to appear.
}

\Ref{
Bernstein, S. (1924).
Sur une modificiation de l'in\'egalit\'e de Tchebichef.
{\it Annals Science Institute Sav.\ Ukraine, Sect.\ Math. I} (in Russian with
French summary).
}

\Ref{
Cirel'son, B.S. (1980).
Quantum generalizations of {Bell's} inequality.
{\it Letters in Mathematical Physics \bf 4}, 93--100.
}

\Ref{
Clauser, J.F., Horne, M.A., Shimony, A., and Holt, R.A. (1969).
Proposed experiment to test local hidden-variable theories.
{\it Phys.\ Rev.\ Letters \bf 49}, 1804--1806.
}

\Ref{
van Dam, W., Gill, R.D. and Grunwald, P. (2002).
Playing games against theories: strengths of nonlocality proofs.
In preparation.
}

\Ref{
Dzhaparidze, K. and van Zanten, H. (2001).
On Bernstein-type inequalities for martingales.
{\it Stochastic Processes and their Applications \bf 93}, 109--117.
}

\Ref{
Einstein, A., Podolsky, B., and Rosen, N. (1935).
Can quantum mechanical description of reality be considered complete?
{\it Phys.\ Rev.\ \bf 47}, 777--780.
}

\Ref{
Feynman, R.P. (1951).
The concept of probability in quantum mechanics.
{\it Proc.\ II Berkeley Symp.\ Math.\ Stat.\ and Prob.},
533--541.
Univ.\ Calif.\  Press, Berkeley.
}

\Ref{
Freedman, D.A. (1975).
On tail probabilities for martingales.
{\it Ann.\ Prob.\  \bf 3}, 100-118.
}

\Ref{
van de Geer, S.A.  (1995).
Exponential inequalities for martingales, with application to maximum likelihood
estimation for counting processes.
{\it Ann.\ Statist.\ \bf 23}, 1779--1801.
}

\Ref{
van de Geer, S.A. (2000).
{\it Empirical Processes in M-estimation}.
Cambridge University Press, Cambridge.
}

\Ref{
Gill, R.D. (1998). Critique of `Elements of quantum probability'.
{\it Quantum Probability Communications \bf 10}, 351--361.
}

\Ref{
Gill, R. D. (2001a).
Asymptotics in quantum statistics.
pp.\ 255--285 in: {\it State of the {A}rt in {P}robability and {S}tatistics, 
{F}estschrift for {W}.{R}. van {Z}wet},
de Gunst, M. C. M.,
Klaassen, C. A. J. and van der Vaart, A. W. (eds),
Lecture Notes--Monograph series {\bf 36},
Institute of Mathematical Statistics, Hayward, Ca.
}

\Ref{
Gill, R. D. (2001b).
Teleportation into quantum statistics.
{\it J. Korean Statist.\ Soc.\ \bf 30}, 291--325.
}

\Ref{
Gill, R.D., Weihs, G, Zeilinger, A. and Zukowski, M. (2002).
No time loophole in Bell's theorem;
the Hess--Philipp model is non-local.
{Proc. Nat. Acad. Sci. USA}, to appear.
{\tt http://arXiv.org/abs/quant-ph/0208187}
}

\Ref{
Hardy, L. (1993).
Nonlocality for two particles without inequalities for almost all entangled states.
{\it Phys.\ Rev.\ Letters \bf 71}, 1665--1668.
}

\Ref{
Hess, K. and Philipp, W. (2001a).
A possible loophole in the theorem of Bell.
{\it Proc. Nat. Acad. Sci. USA \bf 98}, 14224--14227.
}

\Ref{
Hess, K. and Philipp, W. (2001b).
BellÕs theorem and the problem of decidability
between the views of Einstein and Bohr.
{\it Proc. Nat. Acad. Sci. USA \bf 98}, 14228--14233.
}

\Ref{
Hoeffding, W. (1963).
Probability inequalities for sums of bounded random variables.
{\it J. Amer.\ Statist.\ Assoc. \bf 58}, 13--30.
}

\Ref{
{}'t Hooft, G. (1999).
Quantum gravity as a dissipative deterministic system.
{\it Class.\ Quant.\ Grav.\ \bf 16}, 3263-3279.
}

\Ref{
Khrennikov, A.Yu. (1995a).
$p$-adic probability interpretation of Bell's inequality
paradoxes.
{\it Physics Letters A \bf 200}, 119--223.
}

\Ref{
Khrennikov, A.Yu. (1995b).
$p$-adic probability distribution of hidden variables.
{\it Physica A \bf 215}, 577--587.
}

\Ref{
Khrennikov, A.Yu.(1997).
{\it Non-Archimedean analysis: quantum
paradoxes, dynamical systems and biological models.}
Kluwer Acad. Publishers, Dordrecht/Boston/London.
}

\Ref{
Khrennikov A.Yu. (1998).
$p$-adic stochastic hidden variable model.
{\it J. Math.\ Physics \bf 39}, 1388--1402.
}

\Ref{
Khrennikov, A.Yu. (1999)
{\it Interpretations of Probability.}
VSP Int.\ Sc.\ Publishers.
Utrecht, Tokyo.
}

\Ref{
K\"ummerer, B. and Maassen, H. (1998). Elements of quantum probability.
{\it Quantum Probability Communications \bf 10}, 73--100.
}

\Ref{
Larsson, J.-\AA. (2002). A Kochen-Specker inequality. 
{\it Europhys. Lett. \bf 58}, 799--805.
}

\Ref{
Larsson, J.-\AA., and Semitecolos, J. (2001). 
Strict detector-efficiency bounds for $n$-site Clauser-Horne inequalities. 
{\it Phys.\ Rev.\ A \bf 63}, 022117 (5 pp.).
}

\Ref{
Lenglart, E. (1977).
Relation de domination entre deux processus.
{\it Ann.\ Inst.\ Henri Poincar\'e \bf 13}, 171--179.
}

\Ref{
Malley, J. D., and Hornstein, J. (1993).
Quantum statistical inference.
{\it Statistical Science \bf 8}, 433--457.
}

\Ref{
Massar, S. (2001).
Nonlocality, closing the detection loophole and computation complexity.
{\tt http://arXiv.org/abs/quant-ph/0109008}
}

\Ref{
Maudlin, T. (1994).
{\it Quantum Non-locality and Relativity}.
Blackwell, Oxford.
}

\Ref{
Pollard, D. (2001).
{\it User's Guide to Measure Theoretic Probability.}
Cambridge University Press, Cambridge.

}
\Ref{
Pitowsky, I. (1989).
{\it Quantum Probability, Quantum Logic.}
Lecture Notes in Physics {\bf 321}.
Springer-Verlag, Berlin.
}

\Ref{
Rowe, M.A., Kielpinski, D., Meyer, V., Scakett, C.A., Itano, W.M., Monroe, C.,
and Wineland, D.J. (2001).
Experimental violation of a Bell's inequality with efficient detection.
{\it Nature \bf 409}, 791--794.
}

\Ref{
Shorack, G.R. and Wellner, J.A. (1986).
{\it Empirical Processes with Applications in Statistics.}
Wiley, New York.
}

\Ref{
Steiger, W.L. (1969).
{A best possible Kolmogoroff-type inequality for martingales
and a characteristic property.}
{\it Ann.\ Math.\ Statist.\ \bf 40}, 764--769.
}

\Ref{
Tittel, W., Brendel, J., Zbinden, H. , and Gisin, N. (1998).
Violation of Bell inequalities by photons more than 10 km apart.
{\it Physical Review Letters \bf 81},
3563-3566.
}

\Ref{
Thorisson, H. (2000).
{\it Coupling, Stationarity, and Regeneration.}
Springer-Verlag, New York.
}

\Ref{
Weihs, G., Jennewein, T., Simon, C. , Weinfurter, H. and 
Zeilinger, A. (1998).
Violation of {B}ell's inequality under strict {E}instein locality conditions.
{\it Phys. Rev. Lett.  \bf 81}, 5039--5043.
}

\Ref{
Williams, D. (2001).
{\it Weighing the Odds.}
Cambridge University Press,
Cambridge.
}

\Ref{
Whittle, P. (1970).
{\it Probability via Expectation.}
Springer-Verlag, New York. Third edition, 1992.
}


\address
{Mathematical Institute\\
University Utrecht\\
P.O. Box 80010\\
3508 TA Utrecht\\
Netherlands\\}
\email {gill@math.uu.nl}\\
\email{http://www.math.uu.nl/people/gill}

\end{document}